\documentclass[aps,prb,twocolumn,superscriptaddress,showpacs]{revtex4-1}
\usepackage{eurosym}
\usepackage{amsfonts}
\usepackage{amssymb}
\usepackage{amsmath}
\usepackage{graphicx}

\begin{document}

\title{Prediction of superconductivity of $3d$ transition-metal based antiperovskites
via magnetic phase diagram}
\author{D. F. Shao}

\affiliation{Key Laboratory of Materials Physics, Institute of Solid State Physics,
Chinese Academy of Sciences, Hefei 230031, P. R. China}

\author{W. J. Lu}

\email{wjlu@issp.ac.cn}

\author{P. Tong}

\affiliation{Key Laboratory of Materials Physics, Institute of Solid State Physics,
Chinese Academy of Sciences, Hefei 230031, P. R. China}

\author{S. Lin}

\affiliation{Key Laboratory of Materials Physics, Institute of Solid State Physics,
Chinese Academy of Sciences, Hefei 230031, P. R. China}

\author{J. C. Lin}

\affiliation{Key Laboratory of Materials Physics, Institute of Solid State Physics,
Chinese Academy of Sciences, Hefei 230031, P. R. China}

\author{Y. P. Sun}

\email{ypsun@issp.ac.cn}

\affiliation{Key Laboratory of Materials Physics, Institute of Solid State Physics,
Chinese Academy of Sciences, Hefei 230031, P. R. China}

\affiliation{High Magnetic Field Laboratory, Chinese Academy of Sciences, Hefei
230031, P. R. China}

\affiliation{University of Science and Technology of China, Hefei 230026,
P. R. China}

\begin{abstract}
We theoretically studied the electronic structure, magnetic properties,
and lattice dynamics of a series of $3d$ transition-metal antiperovskite
compounds AXM$_{3}$ by density function theory. Based on the Stoner
criterion, we drew the magnetic phase diagram of carbon-based antiperovskites ACM$_{3}$. In the phase diagram,  compounds with non-magnetic ground state but locating
near the ferromagnetic boundary are suggested to yield sizeable electron-phonon
coupling and behave superconductivity. To approve this deduction,
we systematically calculated the phonon spectra and electron-phonon
coupling of   a series of Cr-based antiperovskites ACCr$_{3}$ and ANCr$_{3}$. The results show that AlCCr$_{3}$, GaCCr$_{3}$, and ZnNCr$_{3}$
could be moderate coupling BCS superconductors. The influence of spin fluctuation  on superconductivity are discussed. Furthermore, other  potential superconducting AXM$_{3}$ including some new
Co-base and Fe-based antiperovskite superconductors are predicted from the magnetic phase diagram. 
\end{abstract}


\maketitle

\section{INTRODUCTION}

Since superconductivity (SC) was found by Onnes \cite{onnes_resistance_1911},
researchers have made a lot of efforts to figure out its mechanism. The BCS
theory \cite{BCS1,BCS2,BCS3} pointed out two electrons with opposite spins can pair with
each other via electron-phonon coupling (EPC). Therefore,   magnetism or spin fluctuation (SF) may
break the pair  and be harmful to SC. The discovery of unconventional superconductors such as cuprate oxide superconductors   \cite{bednorz_possible_1986,wu_superconductivity_1987}, iron based superconductors \cite{HCP-iron-nature,kamihara_iron-based_2008}, Sr$_{2}$RuO$_{4}$\cite{maeno_superconductivity_1994}, Na$_{x}$CoO$_{2}$${\cdot}$$y$H$_{2}$O \cite{NaCoO2-SC-found}, etc., challenges the EPC mechanism. For the unconventional superconductors,  SF may play an important role in the superconducting mechanism, which makes researchers   reconsider the
relation between magnetism and SC.

 Superconducting  MgCNi$_{3}$ \cite{he_superconductivity_2001}, containing a high concentration of magnetic element Ni, attracted a lot of attentions on the role of SF.  MgCNi$_{3}$ has a so-called antiperovskite structure, in which Mg atom locates
at the corner, and six Ni atoms at the face-center together with one C atom at the body center form the C-Ni$_{6}$ octahedron. A van Hove singularity locates just below Fermi level ($E_{F}$), leading to  large density of states (DOS) at $E_{F}$ ($N(E_{F})$) \cite{rosner_prl_2001,Singh-Mazin,Shim}.
It makes the compound  strongly exchange-enhanced and unstable toward to
ferromagnetism (FM) upon hole doping \cite{rosner_prl_2001,Singh-Mazin}.
 Experimentally speaking,
due to the high volatility of Mg and the relatively poor reactivity
of C, it is extremely difficult to synthesize stoichiometric MgCNi$_{3}$  \cite{Mollah,Lee-AM,Gordon-singlecrystal}.
The atomic deficiencies in the antiperovskite system may strongly affect the  properties \cite{Sieber-PRB2007,Shao-jap2013}  and lead to some contradictory results in the reported polycrystalline samples \cite{Mollah}.
Recently, the single crystal samples have been prepared \cite{Lee-AM,Gordon-singlecrystal,Lee-jpcm,Pribulova,Diener,Jang,Hong-phonon-singlecrystal}.
The physical property measurements demonstrate the conventional
EPC mechanism, but the reported EPC strengths of these samples are
contradicted \cite{Gordon-singlecrystal,Lee-jpcm,Pribulova,Diener,Jha-mgcni3phonon,tong}.
Moreover,   it has not been successful so far to induce FM instability by preparing
hole-doped compounds such as Mg$_{1-x}$Na$_{x}$CNi$_{3}$ and Mg$_{1-x}$Li$_{x}$CNi$_{3}$.
The interplay between SC and SF in MgCNi$_3$ is still unclear and in debate \cite{Loison-RBPd3}.

Nowadays,  lots of $3d$ transition-metal based antiperovskite
AXM$_{3}$ (A usually is main group element; X = B, C, and N; M = $3d$ transition-metal elements) have
been experimentally synthesized \cite{Fruchart}. However, the reported physical property measurements
are mainly focused on the Ni- and Mn-based antiperovskites. In Ni-based antiperovskites, CdCNi$_{3}$
and ZnNNi$_{3}$  were reported to show SC behavior \cite{CdCNi3,ZnNNi3}.  Abundant
magnetism appears in  Mn-based antiperovskites (see recent review article  Ref. \onlinecite{Tong-AXMn3-review}).
Moreover, superconducting trace has been found in InBSc$_{3}$ \cite{Wiendlocha-InBSc3}.
Some theoretical predictions suggest that SC may exist in some Sc-based
and Cr-based antiperovskites \cite{Wiendlocha-ABSc3,Wiendlocha-ANCr3,RhNCr3,GaNCr3-sc}.
 Indeed, due to the high
concentration of $3d$ transition-metal atoms, one can imagine that
more interesting properties exist in other antiperovskite AXM$_{3}$. The investigation
of them will be very important, both in the search for new superconductors
and in the pursuit of a better understanding of the interplay between
SC and magnetism \cite{Loison-RBPd3}.

In the present work, we offer an approach to explore new $3d$ transition-metal based antiperovskite superconductors. In order to have an overall
understanding of the magnetism in AXM$_{3}$, we calculated the electronic
structures of a series of $3d$ transition-metal based antiperovskites
AXM$_{3}$  and concluded
the doping effects of each atom. From the analysis of these doping effects,
we can evaluate the variation of $N(E_{F})$ of different AXM$_{3}$.
We drew a magnetic phase diagram of ACM$_{3}$ based on the
Stoner criterion $N(E_{F})I>1$. From the phase diagram, we predict  that the materials locating  near FM boundary may have sizeable EPC and could show SC. To confirm this deduction, we systematically  investigated
the formation energies, electronic structures, lattice dynamic
properties, and EPC of a series of Cr-based AXCr$_{3}$. For comparison, such properties of MgCNi$_{3}$ are also calculated. Our results confirm the strong EPC for MgCNi$_{3}$  and suggest that
AlCCr$_{3}$, GaCCr$_{3}$,   and ZnNCr$_{3}$  are moderate coupling BCS superconductors. The depairing effects
from SF are aslo discussed. Furthermore, some  potential superconducting antiperovskites such as new Co-
and Fe-based superconductors are suggested.

\section{METHODS}

The structural relaxation and electronic structure calculations were
performed using projected augmented-wave (PAW) \cite{blochl1994projector,torrent2008implementation}
method  as implemented in the ABINIT code  \cite{gonze2002firstprinciples,gonze2009abinitfirstprinciples,gonze_brief_2005}.
 Electronic wavefunctions are expanded with plane
waves up to an energy cutoff ($E_{cut}$) of 1200 eV. Brillouin zone
sampling is performed on a Monkhorst-Pack (MP) mesh \cite{monkhorst1976special}
of $16\times16\times16.$ The self-consistent calculations were considered
to be converged when the total energy of the system was stable within
$10^{-6}$ Ha. Non-magnetic (NM) and FM states were tested in the study.

The calculations of phonon spectra and EPC were based on the frame work
of the self consistent density functional perturbation theory (DFPT) \cite{DFPT}
using planewaves and ultrasoft pseudopotentials \cite{US} with QUANTUM-ESPRESSO \cite{QE}.
We use an $8\times8\times8$ grid for zone integration in the self-consistent
calculations, while a denser $16\times16\times16$ grid was used in
the EPC calculations. We have calculated dynamical matrices at a
uniform $4\times4\times4$ grid of $q$-points.

To ensure our calculation reliability, we cross-checked the results
given by the above two DFT codes and found them to be in close agreement.
And for consistency, we used the same GGA-PBE \cite{perdew1996generalized}
exchange-correlation potential in both cases.

\section{RESULTS AND DISCUSSIONS}

\subsection{Doping effects of AXM$_{3}$}

As Rosner  \textit{et al.}  \cite{rosner_prl_2001} mentioned, the A-site atom
only plays the role as an effective valent electrons supplier. The
most common A-site atoms of the antiperoveskite compounds can be divided into four groups according to   the effective valence electrons: Cu and Ag (A$^{1+}$); Mg, Zn, and Cd (A$^{2+}$); Al, Ga, and In (A$^{3+}$); Ge and Sn (A$^{4+}$).  Clearly, the effective valence electrons
of A-site atoms are usually in the range from 1 to 4. For simplification,
we chose Na$^{1+}$, Mg$^{2+}$, Al$^{3+}$, and Si$^{4+}$ atoms
as the A-site atoms of the antiperovskites we investigated.

\begin{figure}
 
\includegraphics[width=0.9\columnwidth]{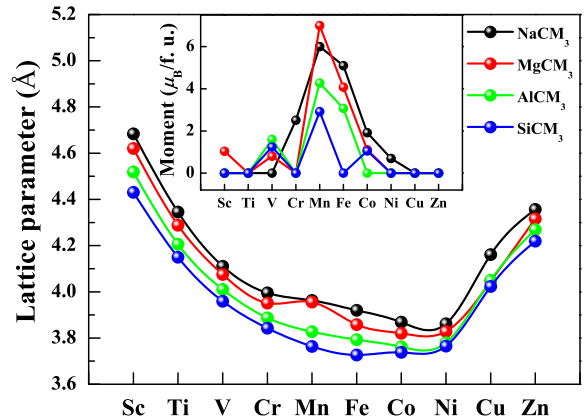}
 
\caption{Lattice parameters and magnetic moments of ACM$_{3}$ (A = Na,
Mg, Al, and Si, M = $3d$ transition metal elements).}
\label{Fig-lattice-moment}

\end{figure}

Figure \ref{Fig-lattice-moment} shows the lattice parameters and magnetic
moment of a series of carbon-based ACM$_{3}$. As
expected, the lattice parameter varies with the same trend of variation
of the atom radius. The ground states of these compounds are determined
by comparing the total energies of the NM and FM states.
The ACMn$_{3}$ and most ACFe$_{3}$ show FM state, which coincides with the experimental
results. Surprisingly, SiCFe$_{3}$ and AlCCo$_{3}$ show NM state.
Comparing the magnetic moments with the $N(E_{F})$ (table \ref{Table-ACM3-N(EF)}),
one can see that all the compounds with FM ground state have quite
large $N(E_{F})$ in NM state. It is corresponding to the Stoner's
itinerant magnetism: the higher $N(E_{F})$ in NM state, the
easier a system becomes spin-polarized. The magnetism of these compounds
will be discussed below.

\begin{table*}
\caption{\label{Table-ACM3-N(EF)}The $N(E_{F})$ (states/eV/spin) in NM ground state (above ``/") and Stoner parameter $I$ (under ``/") of ACM$_{3}$
(A = Na, Mg, Al, and Si, M =  $3d$ transition metal elements). }


\begin{tabular}{ccccccccccc}
\hline
\hline
 & \textbf{M$\rightarrow$}  & \textbf{Sc}  & \textbf{Ti}  & \textbf{V}  & \textbf{Cr}  & \textbf{Mn}  & \textbf{Fe}  & \textbf{Co}  & \textbf{Ni}  & \textbf{Cu}\tabularnewline
\textbf{A$\downarrow$}  &  &  &  &  &  &  &  &  &  & \tabularnewline
\hline
\textbf{Na}  &  & 3.66/0.26 & 3.19/0.22 & 1.51/0.50  & 4.17/0.32  & 7.84/0.37  & 7.61/0.30  & 3.83/0.41 & 7.45/0.33  & 0.44/0.74\tabularnewline
\textbf{Mg}  &  & 5.30/0.30 & 1.64/0.40 & 4.28/0.23 & 2.86/0.34 & 5.47/0.35 & 6.20/0.32  & 3.69/0.41 & 2.57/0.28  & 0.41/0.79\tabularnewline
\textbf{Al}  &  & 0.26/0.29 & 1.491/0.05 & 4.31/0.32  & 2.22/0.32 & 4.87/0.45 & 5.53/0.39 & 0.77/0.74  & 1.13 /0.58  & 1.04/0.20\tabularnewline
\textbf{Si}  &  & 1.26/0.42 & 2.23/0.25 & 6.38/0.21 & 3.06/0.27 & 6.20/0.47 & 1.35/0.31 & 3.87/0.42 & 1.68/0.19 & 0.75/0.72\tabularnewline
\hline
\hline 
\end{tabular}

\end{table*}

\begin{figure}

\includegraphics[width=0.9\columnwidth]{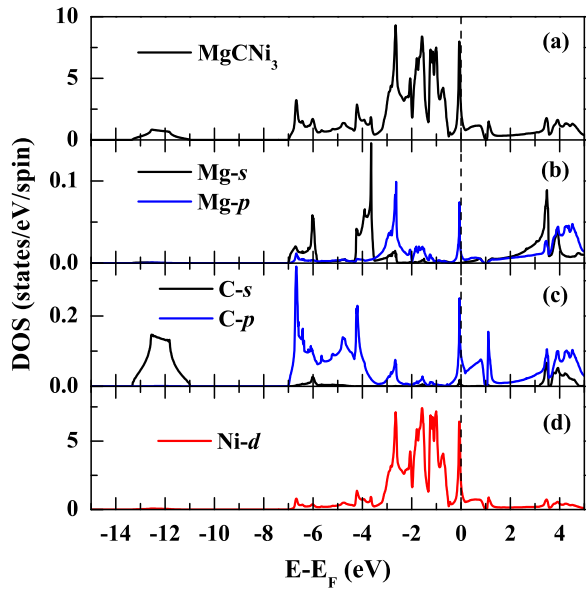}

\caption{Total and atom-orbital-projected DOS of MgCNi$_{3}$.}

\label{Fig-MgCNi3-dos}
\end{figure}

In order to have an overall understanding of the $3d$ transition-metal based antiperovskites AXM$_{3}$, the doping/substitution effects of each
atom on the electronic structure must be investigated. We start
from reviewing the electronic structure of MgCNi$_{3}$. Figure \ref{Fig-MgCNi3-dos}
shows the calculated DOS of MgCNi$_{3}$. It can be seen that the
total DOS near $E_{F}$ are mainly contributed by Ni-$3d$ electrons.
From -7 eV to -4 eV and  near $E_{F}$,   C-$2p$ electrons
hybridize strongly with Ni-$3d$ electrons. The $\pi^{*}$ anti-bonding
state locates just below $E_{F}$, which leads to the  van Hove singularity,
yielding the high $N(E_{F})=2.57$ states/eV/spin. We calculated the
Stoner parameter $I$ using the methods mentioned by Rosner \textit{et al.} \cite{rosner_prl_2001}.
The result shows that $I=0.28$, and Stoner enhancement factor
$S=(1-N(E_{F})I)^{-1}=3.57$,  indicating the strong SF exists
in MgCNi$_{3}$. Our results are very close to the previous theoretical
reports \cite{rosner_prl_2001,Singh-Mazin,Shim,Mollah}, which proves
the validity of present calculations.

\begin{figure}
 
\includegraphics[width=0.9\columnwidth]{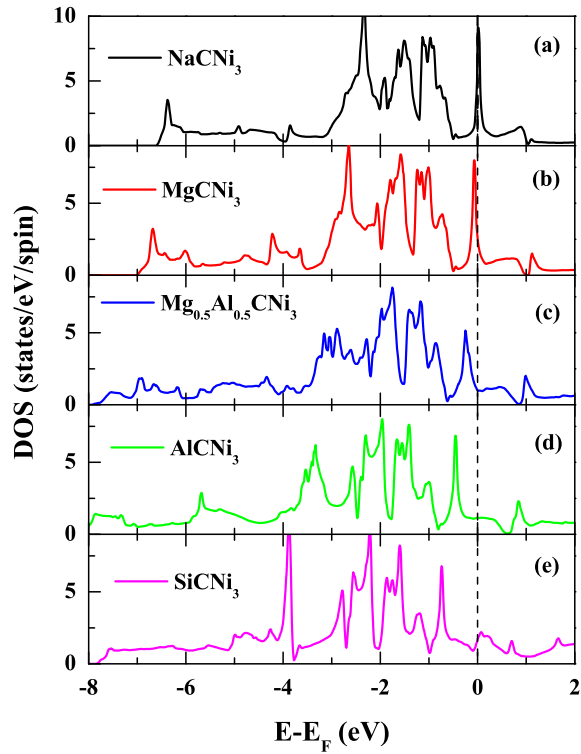}
 
\caption{DOS of (a) NaCNi$_{3}$, (b) MgCNi$_{3}$, (c) Mg$_{0.5}$Al$_{0.5}$CNi$_{3}$,
(d) AlCNi$_{3}$, and (e) SiCNi$_{3}$. }

\label{Fig-ACNi3-dos}
\end{figure}

We firstly studied the A-site doping effect by investigating the electronic
structures of NaCNi$_{3}$, MgCNi$_{3}$, AlCNi$_{3}$, and SiCNi$_{3}$.
Using a doubled supercell, we also calculated the electronic structure
of Mg$_{0.5}$Al$_{0.5}$CNi$_{3}$. As shown in figure \ref{Fig-ACNi3-dos}, the doping of A-site atoms
does not change the overall shape of DOS of ACNi$_{3}$ near $E_{F}$. The
main change is the $E_{F}$ moving with the variation of the amount
of electrons. Thus  the doping effect of A-site atom of AXM$_{3}$
can be evaluated by the rigid band approximation. For NaCNi$_{3}$,
it can be seen as doped MgCNi$_{3}$  by a hole, which causes $E_{F}$
decreasing. The $E_{F}$ just locates at the DOS peak, leading to  very
large $N(E_{F})$, which makes NaCNi$_{3}$ satisfied with the Stoner
criteria $N(E_{F})I>1$. Therefore, spin polarization  appears
in NaCNi$_{3}$.

\begin{figure}
 
\includegraphics[width=0.9\columnwidth]{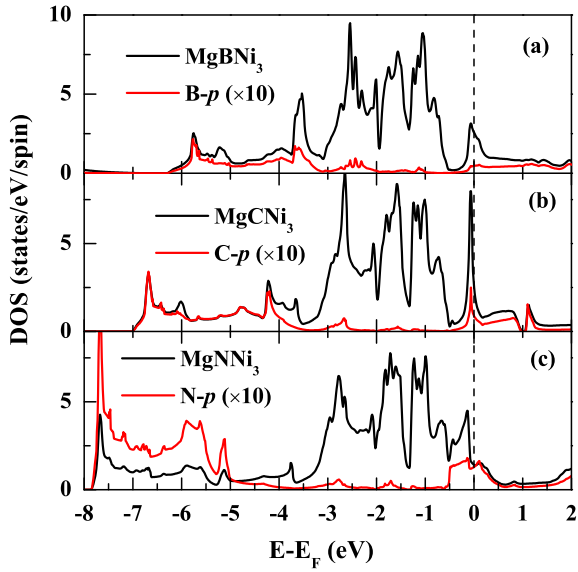}
 
\caption{DOS of (a) MgBNi$_{3}$, (b) MgCNi$_{3}$, and (c) MgNNi$_{3}$. }

\label{Fig-MgXNi3-dos}
\end{figure}

Next we consider the X-site doping effect. Figure \ref{Fig-MgXNi3-dos}
shows the DOS of MgBNi$_{3}$, MgCNi$_{3}$ and MgNNi$_{3}$. Since the X-$2p$
electrons strongly hybridize with Ni-$3d$ electrons,  the change of X-site atoms is expected to strongly influence
the shape of DOS. The evaluation of X-site doping
effect needs to investigate the X-M hybridization besides the $E_{F}$ moving.  For instance, replacing
the carbon atom in MgCNi$_{3}$ by boron atom,  $E_{F}$ moves
to the DOS peak. On the other hand, the B-Ni hybridization makes the peak
smeared, which does not lead to very large $N(E_{F})$. Therefore
MgBNi$_{3}$ is still in the NM ground state. The calculated $N(E_{F})$ of MgBNi$_{3}$
is 2.58 states/eV/spin, which can be comparable to that of MgCNi$_{3}$.
For MgNNi$_{3}$, the DOS peak is smeared, too. Meanwhile $E_{F}$ moves away
from the DOS peak, leading to   small $N(E_{F})=1.36$ states/eV/spin.

\begin{figure}
 
\includegraphics[width=0.9\columnwidth]{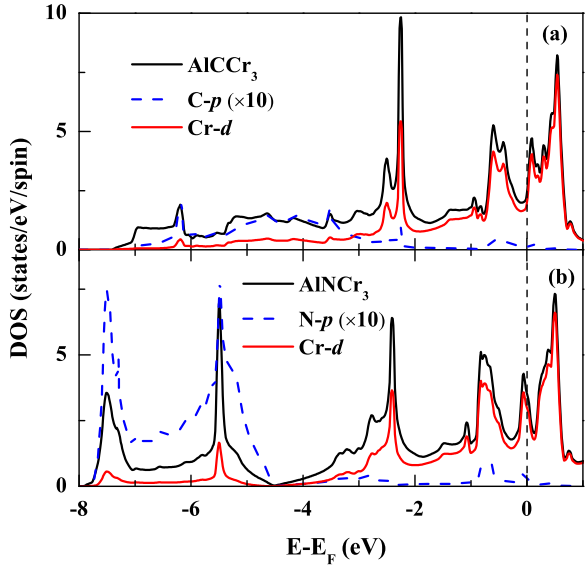}
 
\caption{DOS of (a) AlCCr$_{3}$ and (b) AlNCr$_{3}$.}

\label{Fig-AlXCr3-dos}
\end{figure}
According to our calculation, except for Ni-  and Cu-based AXM$_{3}$,
all the other compounds have similar character that the bonding and
anti-bonding states are far from $E_{F}$. That means for most AXM$_{3}$
compounds, there exists weak X-M hybridizations at $E_{F}$. Figure
\ref{Fig-AlXCr3-dos} shows the DOS of AlCCr$_{3}$ and AlNCr$_{3}$.
The shapes of DOS near $E_{F}$ are very similar. The only difference
is the location of $E_{F}$ decided by total electrons. Thus  for most
AXM$_{3}$ (except for Ni- and Cu-based ones)  the influence of electronic structure by changing X-site
atom can be approximately evaluated as hole or electron doping.

\begin{figure}
 
\includegraphics[width=0.9\columnwidth]{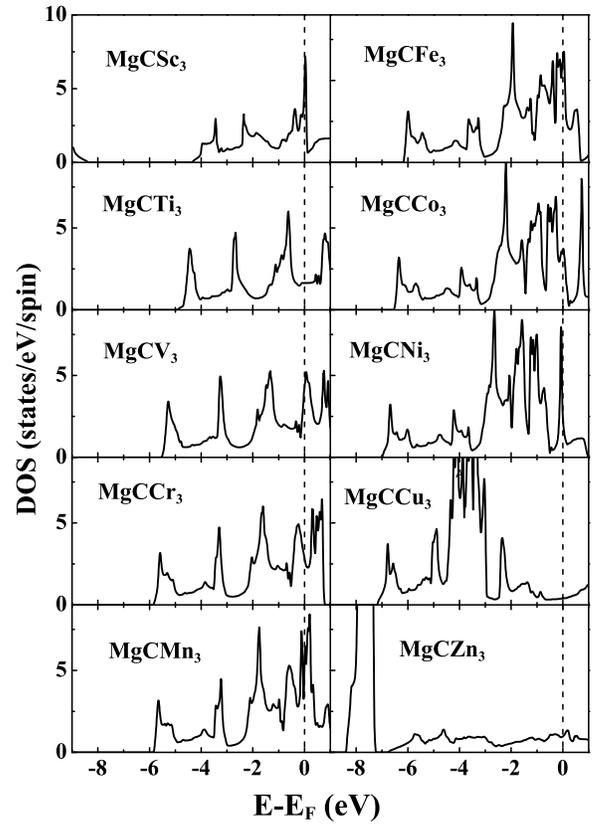}
 
\caption{DOS of MgCM$_{3}$ (M =  $3d$ transition metal elements).}

\label{Fig-MgCM3-dos}
\end{figure}

Now we study the M-site doping effect on the electron structure. Figure \ref{Fig-MgCM3-dos} shows the DOS of MgCM$_{3}$.
The shape of the DOS of MgCZn$_{3}$ is quite different from
those of other MgCM$_{3}$. It is possibly due to that all the
orbitals of Zn are fulfilled, which makes electrons of Zn atom localized.  From MgCSc$_{3}$ to MgCCu$_{3}$, $E_{F}$
moves towards to high energy as total electrons increase, and
the shapes of DOS keep similar feature. Can we just deduce the M-site
doping effect using rigid band approximation? The answer is negative.

\begin{figure}
 
\includegraphics[width=0.9\columnwidth]{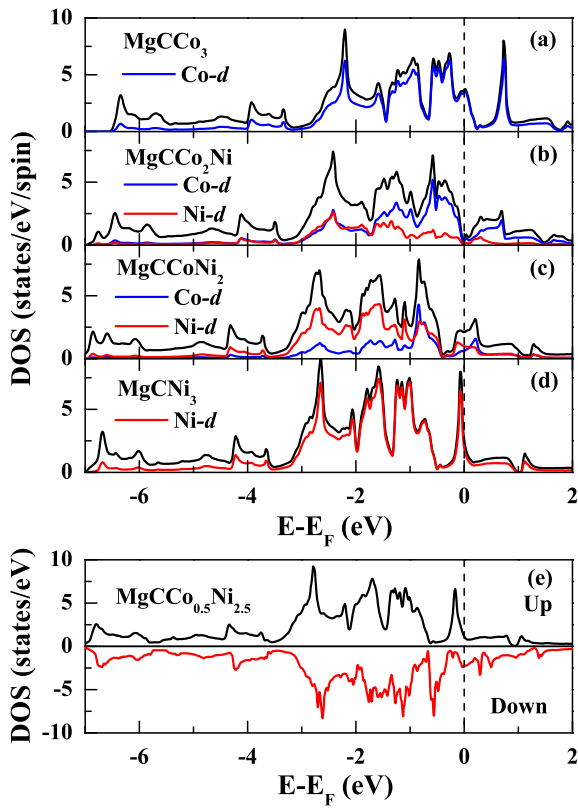}\caption{
\label{Fig-MgCCoNi-dos}DOS in NM state of (a) MgCCo$_{3}$, (b) MgCCo$_{2}$Ni, (c) MgCCoNi$_{2}$,
(d) MgCNi$_{3}$ and (e) DOS in FM state of  MgCCo$_{0.5}$Ni$_{2.5}$. }

\end{figure}

We further investigated the electronic structure of doped MgCNi$_{3}$
by Co (i.e. MgC(Ni,Co)$_{3}$). The calculated DOS in NM state of the doped MgCNi$_{3}$
by one, two, and three cobalt atoms are shown in figures \ref{Fig-MgCCoNi-dos}
(a)-(d). $E_{F}$ of MgCCo$_{3}$ locates at a DOS peak, which makes
MgCCo$_{3}$ be in FM  ground state. For MgCCoNi$_{2}$ and MgCCo$_{2}$Ni, the DOS
near $E_{F}$ can be seen as the sum of the DOS of Ni-$3d$ electrons and Co-$3d$ electrons. The small $N(E_{F})$ of MgCCoNi$_{2}$ and MgCCo$_{2}$Ni lead them to  show NM ground state.
The DOS peak (van Hove singularity) splits into two peaks contributed
by $\pi^{*}$ anti-bonding states of C-Ni and C-Co, respectively.
The Co doping decreases the total electrons, which reduces $E_{F}$
and makes the $\pi^{*}$ anti-bonding state of C-Ni unoccupied. On
the other hand, the Co doping weakens the DOS peak of Ni-$3d$ electrons and enhances the  DOS peak of Co-$3d$ electrons. From the above discussion, one can   deduce that
as MgCNi$_{3}$ is doped by a small Co content, the DOS peak of Ni-$3d$ electrons can still has considerable
strength. Meanwhile $E_{F}$ moves through the peak, which will lead
to   very large $N(E_{F})$. According to Stoner criteria $N(E_{F})I>1$,
the system will become FM state. The calculation of MgCCo$_{0.5}$Ni$_{2.5}$ proves the deduction
(figure \ref{Fig-MgCCoNi-dos} (e)). However such a FM state has not
been observed in experiments \cite{Mollah}, might   due to a high doping
content or an inaccurate stoichiometry. Similarly, the variations
of $N(E_{F})$ and magnetic properties of M-site doping in other
AXM$_{3}$ can be evaluated, too.

\begin{figure}
 
\includegraphics[width=0.9\columnwidth]{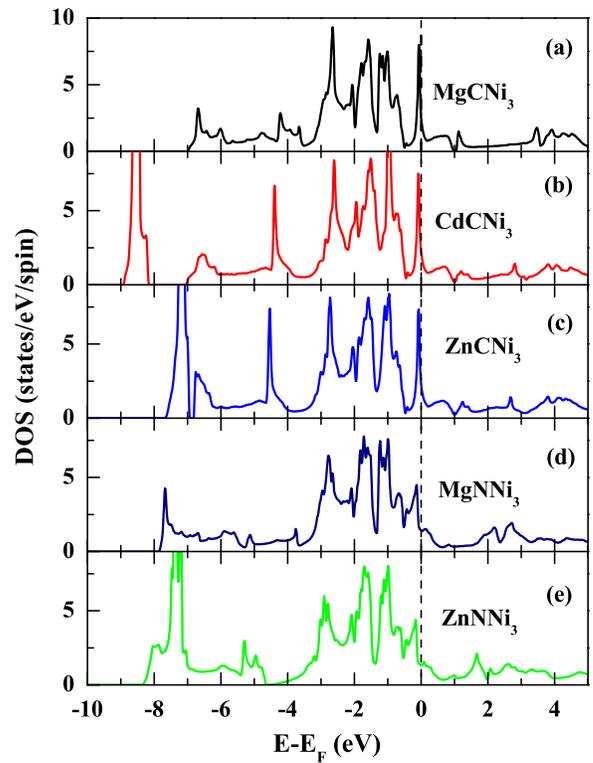}\caption{\label{Fig-AXNi3-dos}DOS of (a) MgCNi$_{3}$, (b) CdCNi$_{3}$, (c) ZnCNi$_{3}$, (d) MgNNi$_{3}$,
and (e) ZnNNi$_{3}$.}

\end{figure}

Additionally, the doping effect of some isovalent A-site atoms was also
investigated. As shown in figure \ref{Fig-AXNi3-dos}, changing Mg to the
isovalent atoms Zn and Cd, the electronic structure near $E_{F}$ rarely
changes. It supports the A-site atoms only give the effective valence
electrons as Rosner \textit{et al.} proposed \cite{rosner_prl_2001}. Thus
the doping effects of A-, X-, and M-sites we concluded above can be generally applied to most $3d$
transition-metal based antiperovskites.

\subsection{Magnetic phase diagram of ACM$_{3}$}

\begin{figure}
 
\includegraphics[width=0.95\columnwidth]{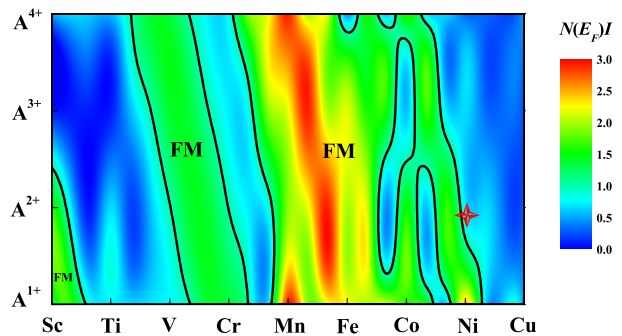} \caption{\label{Fig-phase}Magnetic phase diagram of ACM$_{3}$. The black line is the NM-FM
boundary.  MgCNi$_{3}$ is highlighted using the red star.}

\end{figure}

We calculated the Stoner parameters $I$ of a series of carbon-based ACM$_{3}$ (see table
\ref{Table-ACM3-N(EF)}). According to the doping effects we
 concluded above, the variation of $N(E_{F})$ of different AXM$_{3}$
can be evaluated.  Using the factor $N(E_{F})I$,
we drew the magnetic phase diagram of carbon-based ACM$_{3}$ in figure \ref{Fig-phase}.

The ACM$_{3}$ locating in the areas with the colors green to red are satisfied with the Stoner
criteria $N(E_{F})I>1$ and therefore show FM  ground state. All the Mn-based antiperovskites  and most
Fe-based antiperovskites show FM. Here we must point out that we only consider
the NM and FM states. In reality, abundant magnetisms, e.g. antiferromagnetism
and non-collinear magnetism, are observed experimentally in the Mn-based
antiperovskites \cite{Tong-AXMn3-review}. In this sense, our phase
diagram can not present the real magnetism of such compounds. But
in principle, if an itinerant system has very high $N(E_{F})$ in
NM state, this system is unstable and prefers to be spin-polarized. Accordingly, the magnetic ordering
must occur to lower $N(E_{F})$ to stabilize the system. Therefore,
if a compound locates in the FM area of our phase diagram, it means that
the NM state does not favor the energy minimum  and the magnetic ordering
will emerge. Such phase diagram can help to explore new magnetic
antiperovskites. Furthermore one can also predict the magnetic state of most boron- and nitrogen-based antiperovskites based on the magnetic phase diagram after considering the X-site doping effect as concluded above.

We highlighted MgCNi$_{3}$ using a red star in the phase diagram.
Obviously, MgCNi$_{3}$ is very close to the FM state. One can imagine
that there must be some interplay between FM and SC in MgCNi$_{3}$.
As mentioned above, in order to figure out such interplay, exploring
more MgCNi$_{3}$-like superconducting antiperovskites is necessary.
We assume SC may appear in the area near the NM-FM boundary, since in such
area  the compounds have sizeable $N(E_{F})$ that can lead to a
strong EPC.

\subsection{Superconductivity in MgCNi$_{3}$ and AXCr$_{3}$}

In the magnetic phase diagram, one can notice that some Cr-based
antiperovskites locate in a small interval between two FM areas,
which makes AXCr$_{3}$  as a good system to prove our deduction above.
In this part, we  investigated the potential SC in AXCr$_{3}$. For comparison, the lattice dynamics and EPC property of MgCNi$_{3}$ were also calculated.

Previously, we calculated the formation energies and electronic structures
of a series of Cr-based antiperovskite carbides ACCr$_{3}$ \cite{Shao-ACCr3}.
 Only AlCCr$_{3}$ and GaCCr$_{3}$ have
negative formation energies and may be synthesized experimentally.
Both the two compounds show the NM ground state. Because of the isovalent
A-site atoms, the electronic structures near $E_{F}$ of the two compounds
are almost the same. Similar to MgCNi$_{3}$, $E_{F}$ of the two
compounds locates at a slope of a DOS peak, which leads to  large $N(E_{F})$
and may generate sizeable EPC.

In present work,  we also investigated a series of Cr-based antiperovskite
nitrides ANCr$_{3}$. The lattice parameters and the formation energies
of NM and FM states are listed in table \ref{Table-ANCr3-lattice}.
It can be seen that the magnetic  state of A$^{n+}$NCr$_{3}$
is almost the same as that of A$^{(n+1)+}$CCr$_{3}$, which accords with
the X-site doping effect we concluded above. The calculated results show that ZnNCr$_{3}$, AlNCr$_{3}$,
GaNCr$_{3}$, and SnNCr$_{3}$ have negative formation energies and therefore
may  be  synthesized under normal condition in experiments.

\begin{table*}
\caption{\label{Table-ANCr3-lattice}Lattice parameter $a$,   formation energies $\triangle E$, 
and magnetic moments per Cr atom $M$ of ANCr$_{3}$. }

\begin{tabular}{cccccccccc}
\hline 
\hline 

A & ~~~~~ & {Zn} &{Ga} & {Al} & Ag & Cd & Sn & Mg & In\tabularnewline
\hline 

{$a$ ($\mathrm{\AA}$)}  &  & 3.861 & 3.868 & 3.878 & 3.861 & 3.91 & 3.944 & 3.956 & 3.918\tabularnewline
{$\triangle E_{NM}$} {(eV/atom)}  &  & -0.1027 & -0.2195 & -0.2448 & 0.0972 & 0.0619 & -0.0813 & 0.0173 & 0.0187\tabularnewline
{$\triangle E_{FM}$} {(eV/atom)}  &  & -0.1027 & -0.2206 & -0.2455 & 0.0969 & 0.0626 & -0.0831 & 0.0173 & 0.0186\tabularnewline
{$M$ ($\mu_{B}$/Cr)} &  & 0 & 0.23 & 0.22 & 0.06 & 0 & 0.69 & 0.03 & 0.4\tabularnewline
\hline 
\hline 

\end{tabular}

\end{table*}

\begin{figure}

\includegraphics[width=0.9\columnwidth]{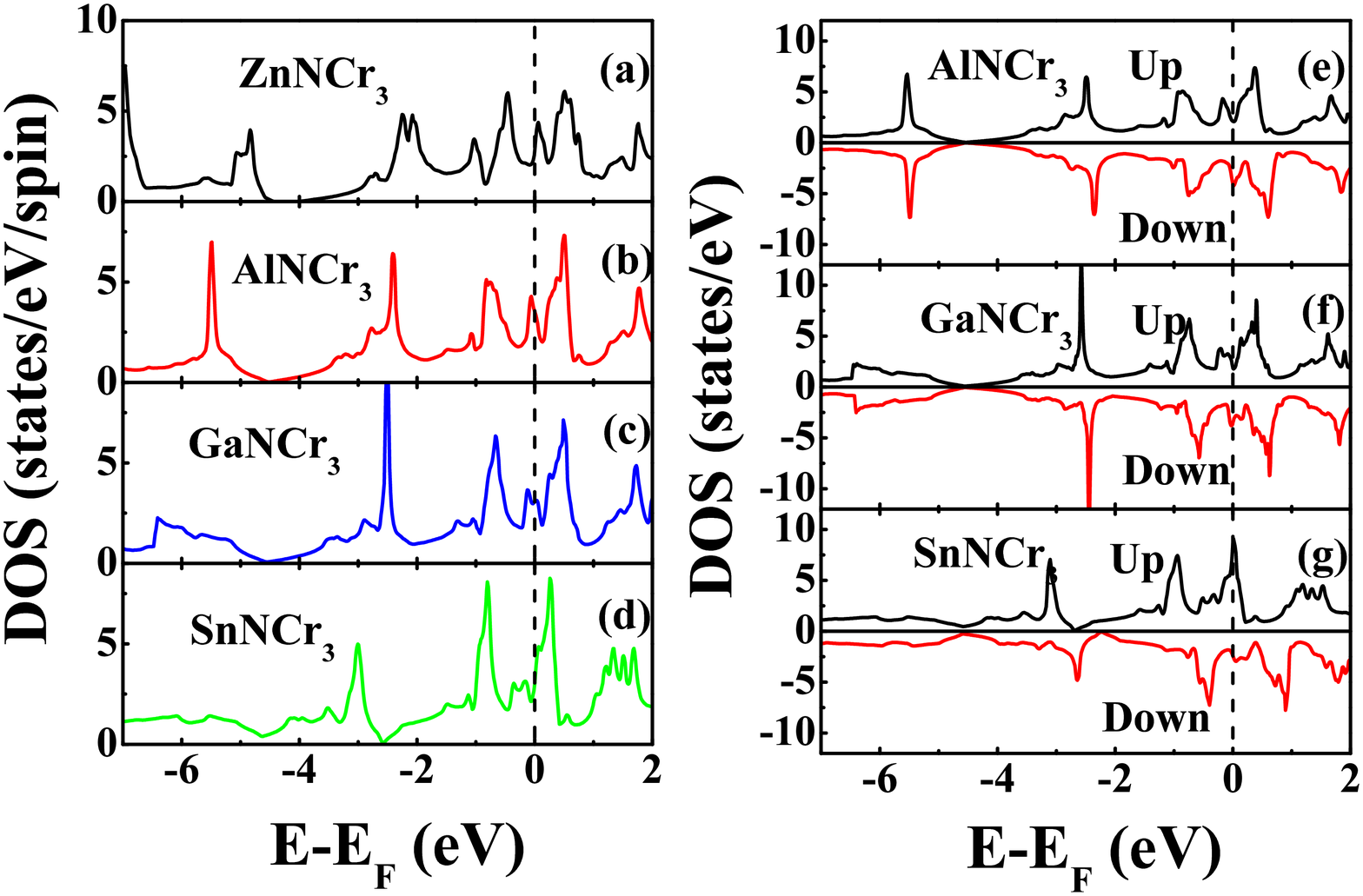}\caption{\label{Fig-ANCr3-dos}Left panel: DOS of (a) ZnNCr$_{3}$, (b) AlNCr$_{3}$,
(c) GaNCr$_{3}$, and (d) SnNCr$_{3}$ in NM state. Right panel: DOS
of (e) AlNCr$_{3}$, (f) GaNCr$_{3}$, and (g) SnNCr$_{3}$ in FM state.}

\end{figure}

The overall shapes of DOS  near $E_{F}$ of ZnNCr$_{3}$, AlNCr$_{3}$,
GaNCr$_{3}$, and SnNCr$_{3}$ are very similar (see figures \ref{Fig-ANCr3-dos}
(a)-(d)). The calculated $N(E_{F})$ and Stoner  criteria $N(E_{F})I$
are listed in table \ref{Table-ANCr3-dos}. It was previously  suggested that
GaNCr$_{3}$ may be a potential superconductor \cite{Wiendlocha-ANCr3,GaNCr3-sc}.
According to  our calculations, it can be seen that $N(E_{F})I\approx1$   for AlNCr$_{3}$ and
GaNCr$_{3}$, which  means the two compounds just locate
at the FM quantum critical point. They show weak FM with small exchange splitting
energies (figures \ref{Fig-ANCr3-dos} (e) and (f)). Therefore we did not consider the possibility of
SC in AlNCr$_{3}$ and GaNCr$_{3}$. It is worth noting that $E_{F}$ of ZnNCr$_{3}$
locates very close to a DOS peak, which leads to  large $N(E_{F})$ and
suggests a sizeable EPC.

\begin{table}
\caption{\label{Table-ANCr3-dos}Calculated $N(E_{F})$ (states/eV/spin) and Stoner  criteria $N(E_{F})I$
of ZnNCr$_{3}$, AlNCr$_{3}$, GaNCr$_{3}$, and SnNCr$_{3}$\textbf{.}}

\begin{tabular}{ccccc}
\hline 
\hline 

 & {ZnNCr$_{3}$}  & {AlNCr$_{3}$}  & {GaNCr$_{3}$}  & {SnNCr$_{3}$}\tabularnewline
\hline 

{$N(E_{F})$ }  & 2.81  & 3.53  & 3.08  & 3.01\tabularnewline
{$N(E_{F})I$}  & 0.76  & 1.09  & 1.01  & 1.29\tabularnewline
\hline 

\end{tabular}

\end{table}

\begin{figure}
\includegraphics[width=0.9\columnwidth]{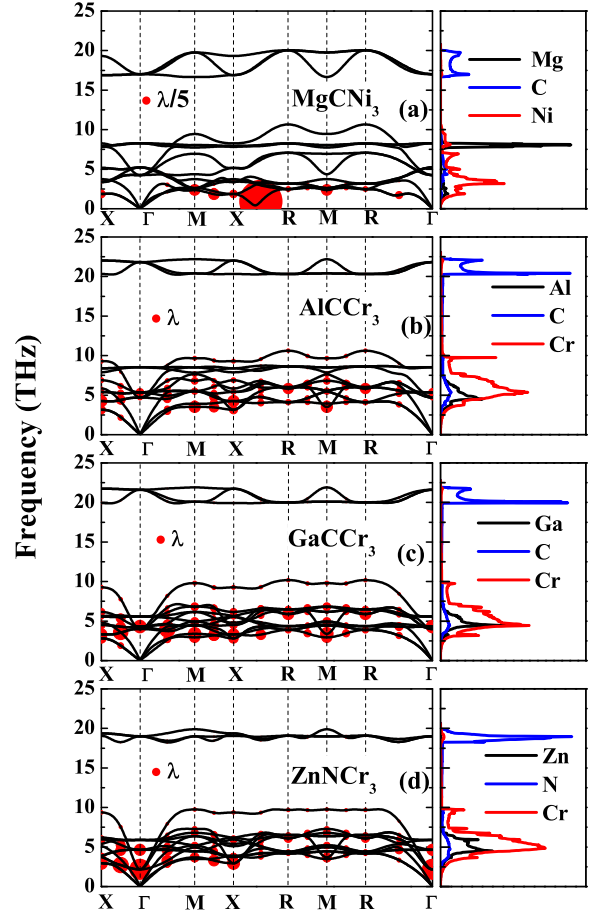}

\caption{Phonon dispersions and phonon DOS of (a) MgCNi$_{3}$, (b) AlCCr$_{3}$,
(c) GaCCr$_{3}$, (d) ZnNCr$_{3}$. The phonon dispersions are decorated
with symbols, proportional to the partial EPC strength $\lambda_{qv}$. }

\label{Fig-phonon}
\end{figure}

We calculated the lattice dynamics and EPC properties
of  MgCNi$_{3}$, AlCCr$_{3}$, GaCCr$_{3}$, and ZnNCr$_{3}$. For reliability, we tested
different exchange-correlation potentials and calculation parameters.
The results are coincide with  each other. The calculated phonon
dispersion curves and the corresponding atom-projected phonon DOS
are shown in figure \ref{Fig-phonon}.  The ideal
antiperovskite in cubic structure (Pm$\overline{3}$m) with five atoms
per unit cell presents fifteen phonon modes including three acoustic and twelve  optical modes. The highest three optical branches resulting mainly from
the lighter C/N atom vibrations are well separated from the other
phonon branches. The phonon modes in low-frequency region come mainly from the vibrations
of Cr/Ni atoms with partial contribution  of  A and C/N vibrations.

\begin{figure*}

\includegraphics[width=0.9\textwidth]{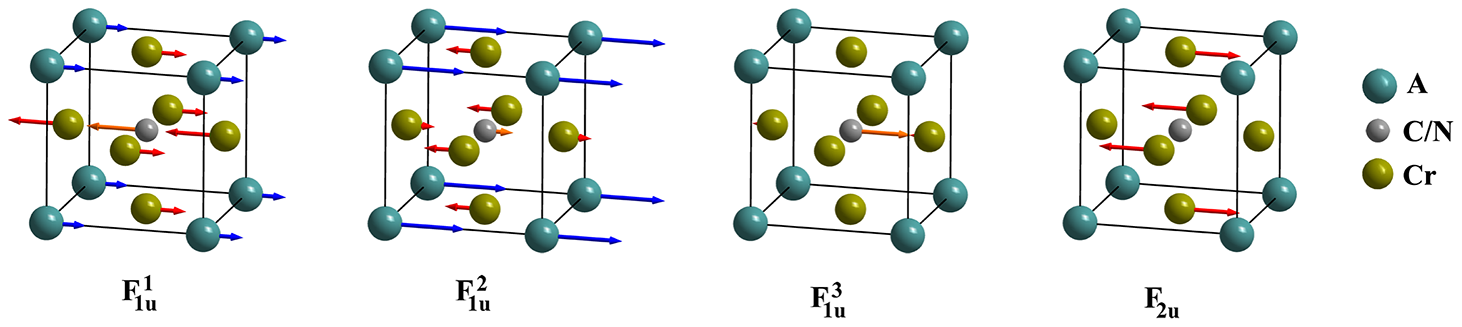}\caption{
\label{Fig-Mode}Schematic eigen displacements of zone-centre optical phonon modes
in AXM$_{3}$. All phonon modes are infrared active.}

\end{figure*}

 Earlier calculations suggested
lattice instabilities existing in MgCNi$_{3}$ \cite{Ignatov,Heid,Jha-PRB,HMTUTU}, which seem to be  ruled out by the recent experimental and theoretical reports \cite{Hong-phonon-singlecrystal,Jha-mgcni3phonon}.
According to our calculation, no imaginary frequency was found in MgCNi$_{3}$,
and the result is similar to the experimental one \cite{Hong-phonon-singlecrystal}.  Soft longitudinal
acoustic (LA) modes at  M and R points are reproduced. Between X  and R point, transverse acoustic (TA) mode of MgCNi$_{3}$
is significantly softened and very close to instability. In previous calculation
by  T\"{u}t\"{u}nc\"{u} \textit{et al.} \cite{HMTUTU}, imaginary frequency exists just
in the same $q$-space path.   Unfortunately, there are no existing experimental
data about the phonon properties in such place. Additionally,
in many previous calculations \cite{Heid,HMTUTU,Ignatov}, strong
softening or imaginary frequency is shown between $\Gamma$   and R points. Our calculation
reproduces such softening. However, experimental
result only shows very small hints of the softening of TA mode   between $\Gamma$   and R points \cite{Hong-phonon-singlecrystal}. The differences between  theoretical and experimental results may be due to the sample quality. As Heid \textit{et al.} \cite{Heid} pointed out, the lattice defects may play an important role in stabilizing the real structure. Indeed, the stoichiometry of the  reported single
crystals are deficient (MgCNi$_{2.8}$ and MgC$_{0.92}$Ni$_{2.88}$),
which may be the reason of the absence of such  soft-phonon anomaly in the
experimental result. Such phonon mode softening is considered as the key role
in the SC mechanism of MgCNi$_{3}$ \cite{Ignatov,walte-rosner,Dolgov}.
In experiments, single crystals with better quality are needed to figure out
such puzzling anomalies.

\begin{table}
\caption{\label{Table-mode}Optical phonon frequencies $\omega$ (THz) at $\Gamma$ point, the calculated EPC strengths $\lambda$, and the logarithmically  averaged phonon frequency
$\omega_{log}$ (K). }

\begin{tabular}{ccccccc}
\hline
\hline 
 & $\omega(F_{1u}^{1})$  & $\omega(F_{1u}^{2})$  & $\omega(F_{1u}^{3}$) & $\omega(F_{2u}$)  & $\lambda$  & $\omega_{log}$ \tabularnewline
\hline 
MgCNi$_{3}$ & 4.26 & 5.21 & 17.01 & 8.28 & 1.34 & 115.77\tabularnewline
AlCCr$_{3}$ & 4.87 & 8.57 & 21.84 & 5.48 & 0.60 & 291.51\tabularnewline
GaCCr$_{3}$  & 4.19 & 5.62 & 21.63 & 4.44 & 0.78 & 253.91\tabularnewline
ZnNCr$_{3}$ & 2.45 & 5.92 & 18.98 & 4.82 & 0.67 & 260.27\tabularnewline
\hline
\hline
\end{tabular}

\end{table}

For AlCCr$_{3}$, GaCCr$_{3}$, and ZnNCr$_{3}$, the absence of imaginary
frequency suggests the dynamic stability of the considered crystal
structures of the three compounds. The acoustic modes are harder than
those of MgCNi$_{3}$, and no modes are close to instability. The optical
phonons at $\Gamma$ point   belong
to the following irreducible representations: $F(O_{h})=3F_{1u}+F_{2u}$
(see figure \ref{Fig-Mode}), which are infrared (IR) active  (table \ref{Table-mode}). The C/N related $F_{1u}^{1}$ mode
and $F_{1u}^{3}$ mode of ZnNCr$_{3}$ are softer than those of
AlCCr$_{3}$ and GaCCr$_{3}$ maybe due to the larger atomic
mass of nitrogen. We did not observe the significant softening of
$F_{2u}$ mode that reported by  T\"{u}t\"{u}nc\"{u} \textit{et al.}  in RhNCr$_{3}$ and GaNCr$_{3}$ \cite{RhNCr3,GaNCr3-sc}.

\begin{figure}

\includegraphics[width=0.9\columnwidth]{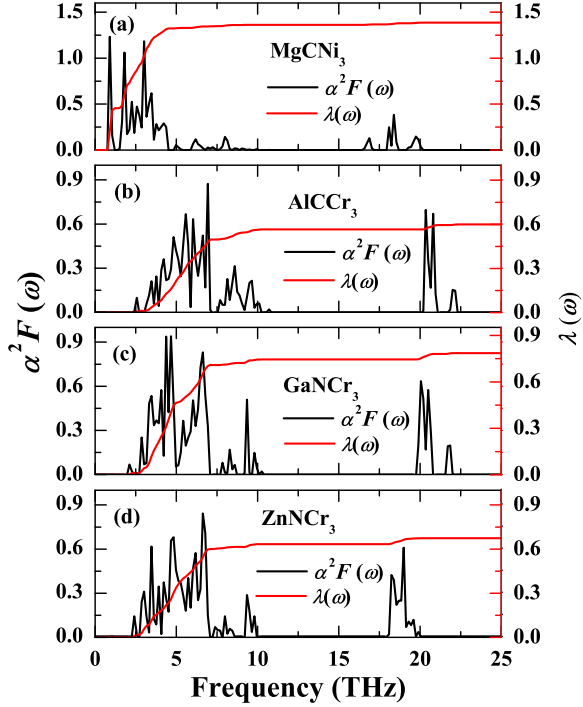}\caption{\label{Fig-a2f}  Eliashberg functions (left) and frequency-dependent EPC strength
$\lambda$ (right) of (a) MgCNi$_{3}$, (b) AlCCr$_{3}$, (c) GaCCr$_{3}$,
and (d) ZnNCr$_{3}$.}

\end{figure}

The calculated Eliashberg functions $\alpha^{2}F(\omega)$ of MgCNi$_{3}$, AlCCr$_{3}$, GaCCr$_{3}$,
and ZnNCr$_{3}$ are plotted in figure \ref{Fig-a2f}  according to the below formula:
\begin{equation}
\alpha^2 F(\omega)=\frac{1}{N(0)}\sum_{\mathbf{k},\mathbf{q},\nu,n,m}%
\delta(\epsilon_{\mathbf{k}}^{n})\delta(\epsilon_{\mathbf{k+q}}^{m}%
)|g_{\mathbf{k},\mathbf{k+q}}^{\nu,n,m}|^{2}\delta(\omega-\omega^{\nu}
_{\mathbf{q}}),
\label{eq:alpha}
\end{equation}
where $\omega^{\nu}_{\mathbf{q}}$ are phonon frequencies,
$\epsilon_{\mathbf{k}}^{n}$ electronic energies, 
and
$g_{\mathbf{k},\mathbf{k+q}}^{\nu,n,m}$  $EP$ matrix
elements. The total EPC strength is
\begin{equation}
\lambda\!=\!\sum_{\mathbf{q},\nu} 
\lambda_{\mathbf{q}}^{\nu}=2 \int_{0}^{\infty} 
\frac{\alpha^2 F(\omega)}{\omega} d \omega.\label{eq:lambda}
\end{equation}
The calculated $\lambda_{\mathbf{q}}^{\nu}$ are visualized as red circles in figure \ref{Fig-phonon}.
For MgCNi$_{3}$, the absence of imaginary
frequency  allows the direct calculation of  the quantitative
EPC contribution details from the lowest acoustic mode. Obviously for MgCNi$_{3}$ the  EPC
originates mainly from acoustic phonon modes softening. For  AlCCr$_{3}$, GaCCr$_{3}$, and ZnNCr$_{3}$, optical modes are playing an important role. The   EPC  distributes
throughout some low-frequency phonon modes. One can note that a larger
 $\lambda_{\mathbf{q}}^{\nu}$ always exists at the frequencies at which    large
Cr-phonon DOS present. At the zone center, the $F_{1u}^{1}$ and $F_{2u}$ modes
mainly contribute to   EPC. For the two modes, Cr atoms vibrate strongly
against each other (figure \ref{Fig-Mode}). Such vibrations can make large
change in the overlap of $3d$ orbitals between neighbouring Cr atoms,
leading to strong coupling. Comparing the EPC strength from different phonon modes of  GaCCr$_{3}$ and ZnNCr$_{3}$, it can be found that  the softening of $F_{1u}^{1}$
mode can strongly enhance
$\lambda$.

 The integrated EPC strengths $\lambda(\omega)$
are calculated using equation (\ref{eq:lambda}), and are   plotted in
figure \ref{Fig-a2f}. The EPC comes mainly from the $\alpha^{2}F(\omega)$
peaks at low frequencies, which are mostly related to the vibrations of Ni/Cr
atoms. For the peaks at high frequencies related to the vibrations of C/N atoms, their contributions to
$\lambda$ are less than 6\%.

The calculated $\lambda$ of MgCNi$_{3}$ is 1.34, implying very
strong EPC. However, the calculated $\lambda$ of AlCCr$_{3}$,
GaCCr$_{3}$, and ZnNCr$_{3}$ are less than 1.0 and about $0.6\sim0.8$ (table \ref{Table-mode}),
implying moderate EPC. The difference of $\lambda$ for AXCr$_{3}$
can be described qualitatively using Hopfield expression
\begin{equation}
\lambda=\frac{N(E_{F})I^{2}}{M\left\langle \omega^{2}\right\rangle },\label{eq:Hopfield-1}
\end{equation}
where  $I$ is mean square EPC matrix element averaged
over Fermi surface,   $M$ is the ion mass, and $\left\langle \omega^{2}\right\rangle $
is the average squared phonon frequency. The higher $N(E_{F})$ and the lower phonon frequency make GaCCr$_{3}$
has the larger $\lambda$.

\begin{figure}

\includegraphics[width=0.9\columnwidth]{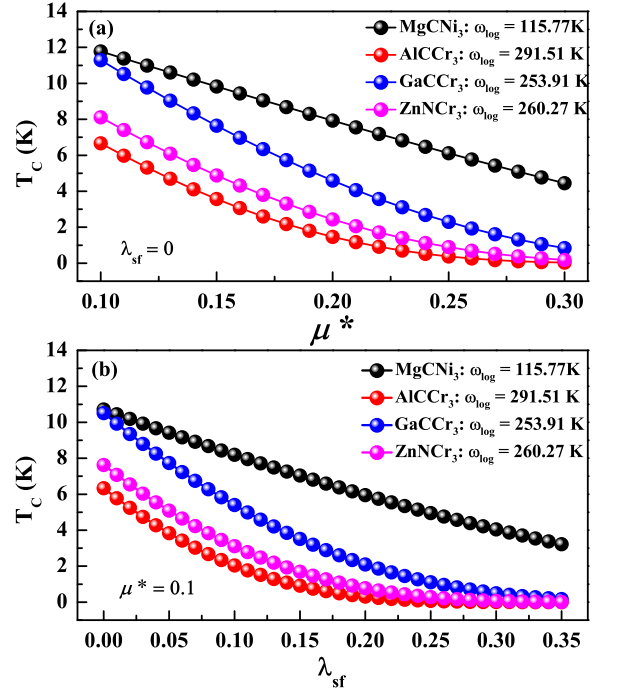}

\caption{\label{Fig-TC}(a) Variation of $T_{C}$ with $\mu^{*}$(purely static pair breaking); (b) Variation of $T_{C}$ with
$\lambda_{sf}$ (electron-paramagnon coupling).}

\end{figure}

To calculate $T_{C}$, we used the Allen-Dynes equation \cite{mcmillan1968transition,Dynes,Allen-dynes}
\begin{equation}
T_c= \frac{\omega_{\log}}{1.20}\exp\left( - \frac{1.04(1+\lambda)}{\lambda-\mu^*-0.62\lambda\mu^*}\right),\label{eq:TC}
\end{equation}
 where $\mu^{*}$ is Coulomb pseudopotential and the logarithmically averaged characteristic phonon frequency $\omega_{log}$
is defined as
\begin{equation}
\omega_{log}=\exp\left(\frac{2}{\lambda}\int\frac{d\omega}{\omega}\alpha^{2}F(\omega)\log\omega\right).
\end{equation}
At the absence of SF, using a typical $\mu^{*}=0.1$, we can get $T_{C}$
with the values of 11.82, 6.67, 11.29, and 8.23 K for MgCNi$_{3}$,
AlCCr$_{3}$, GaCCr$_{3}$, and ZnNCr$_{3}$, respectively.
According to the experimental results of MgCNi$_{3}$ single crystals \cite{Lee-AM,Lee-jpcm,Pribulova,Diener,Jang,Gordon-singlecrystal,Hong-phonon-singlecrystal},
the observed $T_{C}$ is around 7 K. It seems that our calculation overestimates $T_{C}$ of MgCNi$_{3}$. Therefore we need to consider the influence
of SF    in the system.

We firstly considered the purely static pair breaking due to SF by
varying $\mu^{*}$. In figure \ref{Fig-TC}
(a), one can note that the increasing $\mu^{*}$ significantly depresses
$T_{C}$. As $T_{C}$ around 7 K, we can get $\mu^{*}\approx0.23$.
Such a value is smaller than the previous derivations of $\mu^{*}=0.33$ \cite{Ignatov}
and $\mu^{*}=0.41$ \cite{walte-rosner}, but  still larger than
the typical values $\mu^{*}=0.1\sim0.15$ for the absence of SF.

It is better to discuss the influence of SF in terms of electron-paramagnon
coupling strength $\lambda_{sf}$ \cite{Ignatov}. In that case, the
effective coupling strength should be $\lambda_{\Delta}=\lambda-\lambda_{sf}$.
 The effective mass of the carriers should be enhanced by the factor
$1+\lambda_{Z}=1+\lambda+\lambda_{sf}$. And equation (\ref{eq:TC}) shound
be changed as \cite{Dolgov,Oritenzi-Boeri}
\begin{equation}
T_{C}=\frac{\omega_{log}}{1.45}\exp\left(-\frac{(1+\lambda_{Z})}{\lambda_{\Delta}-\mu^{*}(1+0.5\frac{\lambda_{\Delta}}{1+\lambda_{Z}})}\right).\label{eq:Tc-sf}
\end{equation}
Figure \ref{Fig-TC} (b) shows the calculated $T_{C}$ using equation (\ref{eq:Tc-sf})
with different $\lambda_{sf}$ (here the Coulomb pseudopotential
is fixed to the typical value $\mu^{*}=0.1$). When
$T_{C}$ is around 7 K, we can get $\lambda_{sf}\approx0.15$. Although the obtained parameter is small, it still implies that SC in the system is partially suppressed by depairing effect from SF \cite{walte-rosner,Dolgov}. The influence of SF on $T_{C}$ of AlCCr$_{3}$, GaCCr$_{3}$, and ZnNCr$_{3}$
is also present in figure \ref{Fig-TC}. If we consider the same SF influence as in MgCNi$_{3}$, using equation (\ref{eq:TC}) with $\mu^{*}=0.23$ and $\lambda_{sf}=0$,
we can get $T_{C}=0.70$, 3.11, and 1.40 K for AlCCr$_{3}$, GaCCr$_{3}$,
and ZnNCr$_{3}$, respectively; using equation (\ref{eq:Tc-sf}) with $\mu^{*}=0.1$ and $\lambda_{sf}=0.15$,
we can get $T_{C}=0.91$, 3.52, and 1.69 K for AlCCr$_{3}$, GaCCr$_{3}$,
and ZnNCr$_{3}$, respectively.

Let us look back at the electronic structures of AlCCr$_{3}$, GaCCr$_{3}$,
and ZnNCr$_{3}$. The $E_{F}$ locates between two DOS peaks, which suggests
in the system both hole and electron doping can enhance the SF
and lead to a transition from SC to magnetic instability. Considering it
has not been successful so far to investigate such transition in doped
antiperovskites such as Mg$_{1-x}$(A$^{1+}$)$_{x}$CNi$_{3}$ due to the difficulty in experimental preparation of samples, AXCr$_{3}$ may be good
candidates to figure out the relation between SC and magnetism in
antiperovskites.

\subsection{Other potential superconducting AXM$_{3}$}

Since the discovery of SC in MgCNi$_{3}$, researchers have made great
efforts to explore other $3d$ transition-metal based antipervoskite
superconductor \cite{Wiendlocha-InBSc3,Wiendlocha-ABSc3,Wiendlocha-ANCr3,RhNCr3,GaNCr3-sc,Schaak}.
The observed superconducting trace at 4.5 K found in In$_{1.3}$B$_{0.7}$Sc$_{3}$
is encouraging \cite{Wiendlocha-InBSc3}. Theoretically, InBSc$_{3}$
is close to weak FM  and with an EPC superconducting mechanism \cite{Wiendlocha-InBSc3,Wiendlocha-ABSc3}.
One should believe   the Ni-based  antiperovskite superconductor are not alone and other superconducting antiperovskites may be hidden behind them.

\begin{figure}

\includegraphics[width=0.9\columnwidth]{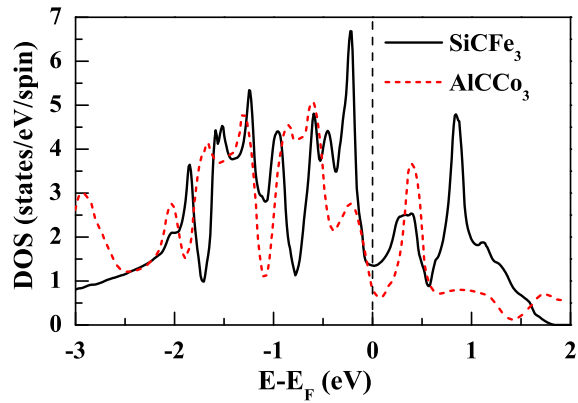}

\caption{DOS near $E_{F}$ of (a) SiCFe$_{3}$  and (b) AlCCo$_{3}$. }

\label{Fig-ACFeCo}
\end{figure}

Indeed, our magnetic phase diagram shows that some AXSc$_{3}$ should be very
close to FM, which may potentially be superconducting. There are also some
other antiperovskite compounds  close to FM, such as AXV$_{3}$, for which the cubic
phase has not been successfully prepared so far. Here we present two
potential superconducting parent materials A$^{4+}$CFe$_{3}$ (A = Si, Ge, etc.) and
A$^{3+}$CCo$_{3}$ (A = Al, Ga, etc.). The DOS near $E_{F}$ of SiCFe$_{3}$ and AlCCo$_{3}$ are
plotted in figure \ref{Fig-ACFeCo}. One can notice that the $E_{F}$ of
the two compounds  just locates at a DOS valley. Thus both electron and
hole doping can move $E_{F}$ to the DOS peaks nearby, and largely
enhance $N(E_{F})$ to yield strong EPC. Therefore, doped A$^{3+}$CCo$_{3}$
may  be  new Co-based superconductors besides Na$_{x}$CoO$_{2}$${\cdot}$$y$H$_{2}$O \cite{NaCoO2-SC-found}.
And doped A$^{4+}$CFe$_{3}$ may be  new Fe-based superconductors.

 As we
know, the role of EPC in the Fe-based superconductors is still puzzling. For instance, EPC mechanism can get a reasonable $T_{C}$ of the hcp iron under
pressure, but can not explain the rapid disappearance of SC above
30 GPa \cite{HCP-iron-nature,mazin-hcpiron,Bose-hcpiron}. For the iron pnictide superconductors \cite{kamihara_iron-based_2008},
Boeri \textit{et al.} \cite{Boeri-PRL2008,Boeri-Physica C2009} and Mazin \textit{et
al.} \cite{Mazin-PRL2008} pointed out that the EPC in the system is
intrinsically weak, but can be enhanced strongly by magnetism \cite{Boeri-Mazin-PRB2008}.
Although EPC is still not enough to explain the high critical temperature,
 it is strong enough to have a non-negligible effect on SC. On the other
hand, in the similar system of nickel pnictide superconductors, it seems
the EPC plays the main role \cite{Boeri-Physica C2009}. One may
connect nickel pnictide superconductors with MgCNi$_{3}$ by the
similar EPC mechanism. The potential supeconductivity in doped A$^{4+}$CFe$_{3}$ may also help to figure out the superconducting
mechanism, which needs to be futher confirmed in experimental and theoretical works.

\section{CONCLUTIONS}

In this work, we theoretically investigated the electronic structure, magnetic properties, and lattice dynamics of the $3d$ transition-metal based antiperovskite compounds AXM$_{3}$. Based on the
analysis of the doping effect in AXM$_{3}$, we drew the magnetic
phase diagram of ACM$_{3}$.
We suggested that superconducting antiperovskites can be found
in the NM area but very close to FM boundary.  In order to prove the deduction, we investigated
a series of Cr-based antiperovskite AXCr$_{3}$ (X=C, N).
 The results
indicate that AlCCr$_{3}$,  GaCCr$_{3}$, and ZnNCr$_{3}$ are of NM ground
state but very close to FM instability. Due to the large $N(E_{F})$, these compounds  have sizeable EPC. We calculated the phonon spectra  and EPC of MgCNi$_{3}$,  AlCCr$_{3}$, GaCCr$_{3}$, and ZnNCr$_{3}$.  Our results confirm the strong EPC in MgCNi$_{3}$, and show that  AlCCr$_{3}$, GaCCr$_{3}$, and ZnNCr$_{3}$
are moderate coupling BCS superconductors.   Such   compounds may be good
candidates to figure out the role of SF.
Moreover, some potential new antiperovskite Co-based and Fe-based superconductors
are suggested.

Up to date, there is no definite evidence of SF in MgCNi$_{3}$ single crystals \cite{Lee-jpcm}.
Since such single crystals are not stoichiometric, the interplay between
SC and magnetism are still puzzling. Our phase diagram
and predictions could offer new route to figure out the issue. In experiments one
can try to prepare such potential antiperovskite superconductors we predicted. Furthermore, by doping we can control the physical properties and may observe the magnetic-superconducting transition through quantum critical point. We hope our phase diagram and
predictions can help to explore more $3d$ transition-metal based
antiperovskite superconductors  and figure out the relation between
SC and magnetism in the system.

\begin{acknowledgments}
This work was supported by the National Key Basic Research under Contract No. 2011CBA00111, and the National Nature Science Foundation of China under Contract Nos. 51171177, 11304320, 11274311, 11174295, and U1232139. The calculations were partially performed at the Center for Computational Science, CASHIPS.
\end{acknowledgments}


\begin{thebibliography}{10}
\bibitem{onnes_resistance_1911} Onnes H K 1911 \emph{Comm. Phys. Lab. Univ. Leiden}  {\bf 122} 124

\bibitem{BCS1}  Cooper L N 1956 \emph{Phys. Rev.} {\bf 104} 1189

\bibitem{BCS2}  Bardeen J,  Cooper L N, and  Schrieffer J R 1957 \emph{Phys.
Rev.} {\bf106} 162 

\bibitem{BCS3} Bardeen J,  Cooper L N, and  Schrieffer J R 1957 \emph{Phys.
Rev.} {\bf108} 1175 




\bibitem{bednorz_possible_1986} Bednorz J  G  and  M{\"u}ller K  A  1986
\emph{Z. Phys. B: Condens. Matter} {\bf64}  189  

\bibitem{wu_superconductivity_1987} Wu M  K,    Ashburn J R,  
Torng C  J,   Hor P  H, and Meng  R L 1987 \emph{Phys. Rev. Lett.} {\bf58}  908 


\bibitem{HCP-iron-nature}Shimizu K,  Kimura T, Furomoto S,  Takeda K,
 Kontani K,  Onuki Y, and  Amaya K 2001 \emph{Nature} {\bf412}  316  



\bibitem{kamihara_iron-based_2008}  Kamihara Y,  Watanabe T,  Hirano M,
and  Hosono H 2008 \emph{J. Am. Chem. Soc.} {\bf130}  3296  

\bibitem{maeno_superconductivity_1994} Maeno Y,  Hashimoto H, 
Yoshida K,  Nishizaki S,  Fujita T,  Bednorz J  G, and  Lichtenberg F 1994
\emph{Nature} {\bf372}  532  


\bibitem{NaCoO2-SC-found}  Takada K,  Sakurai H,  Takayama-Muromachi E,
 Izumi F,  Dilanian R  A, and  Sasaki T 2003  \emph{Nature} {\bf422} 53  

\bibitem{he_superconductivity_2001}  He T,  Huang Q,  Ramirez A P,
 Wang Y,  Regan K A,  Rogado N,  Hayward M A,  Haas M K, 
Slusky J S,  Inumara K,  Zandbergen H W,  Ong N P, and  Cava R J 2001 \emph{Nature}
{\bf411}  54  


\bibitem{rosner_prl_2001}  Rosner H,  Weht R,  Johannes  M D, Pickett W
E, and Tosatti E 2001 \emph{Phys. Rev. Lett.} {\bf88}, 027001  

\bibitem{Singh-Mazin}  Singh D  J  and  Mazin I  I 2001 \emph{Phys. Rev. B} {\bf64} 
140507(R)  

\bibitem{Shim}  Shim J H,  Kwon S K, and  Min B I 2001 \emph{Phys. Rev.
B} {\bf64}  180510(R) 

\bibitem{Mollah}  Mollah S 2004 \emph{J. Phys.: Condens. Matter} {\bf16} R1237 

\bibitem{Lee-AM}  Lee H S，   Jang D J,  Lee H G,  Lee S I, 
Choi S M, and Kim C J 2007  \emph{Adv. Mater.} {\bf19}  1807 

\bibitem{Gordon-singlecrystal} Gordon R  T, Zhigadlo N D,  Weyeneth S,
 Katrych S, and  Prozorov R 2013 \emph{Phys. Rev. B} {\bf87}  094520  

\bibitem{Sieber-PRB2007}  Sieberer M,  Mohn P, and  Redinger J 2007 \emph{Phys.
Rev. B} {\bf75}   024431 

\bibitem{Shao-jap2013}  Shao D  F, Lu W  J,  Lin J C,  Tong P,
Jian H  B, and  Sun Y P 2013 \emph{J. Appl. Phys.} {\bf113}  023905  


\bibitem{Lee-jpcm}  Lee H S,  Jang D J,  Lee H G,  Kang W, Cho  M
H, and  Lee S I 2008 \emph{J. Phys.: Condens. Matter} {\bf20}, 255222  

\bibitem{Pribulova}  Pribulov\'{a} Z,  Ka\v{c}mar\v{c}\'{i}k J,  Marcenat C,  Szab\'{o} P,  Klein T,
 Demuer A,  Rodiere P,  Jang D J, Lee  H S,  Lee S I, and  Samuely P 2011
\emph{Phys. Rev. B} {\bf83}  104511  

\bibitem{Diener}  Diener P,  Rodi\`{e}re P,  Klein T,  Marcenat C, 
Kacmarck J,  Pribulova Z,  Jang D J,  Lee H S,  Lee H G, and 
Lee S I 2009 \emph{Phys. Rev. B} {\bf79}  220508(R)  

\bibitem{Jang}  Jang D J,  Lee H S,  Lee H G,  Cho M H, and 
Lee S I 2009 \emph{Phys. Rev. Lett.} {\bf103} 047003 


\bibitem{Hong-phonon-singlecrystal} Hong H,  Upton M,  Said A
H,  Lee H  S,  Jang D J,  Lee S I,  Xu R, and
Chiang T C  2010 \emph{Phys. Rev. B} {\bf82} 134535 (2010).

\bibitem{Jha-mgcni3phonon}  Jha P K,  Gupta S D, and
 Gupta S K 2012 \emph{AIP Advances} {\bf2} 022120 

\bibitem{tong}  Tong  P and  Sun Y P  2012 \emph{Adv. Cond. Matter Phys.} {\bf2012}
903239

\bibitem{Loison-RBPd3}  Loison C,  Leithe-Jasper A, and  Rosner H 2007 \emph{Phys. Rev. B} {\bf75} 205135

\bibitem{Fruchart}  Fruchart D,  Fruchart R,  L'heritier P,  Kanematsu K,
 Madar R,  Misawa S,  Nakamura Y,  Webster P J, and  Ziebeck K R A 1988
\emph{Alloys and Compounds of D-Elements With Main Group Elements} (Berlin: Springer)

\bibitem{CdCNi3}  Uehara M,  Yamazaki T,  K\^{o}ri T,  Kashida T, 
Kimishima Y, and  Hase I 2007 J. \emph{Phys. Soc. Jpn.} {\bf76} 034714

\bibitem{ZnNNi3} Uehara  M,  Uehara A,  Kozawa K, Yamazaki  T, and
 Kimishima Y 2010 \emph{Physica C} {\bf470}S688  

\bibitem{Tong-AXMn3-review}  Tong P, Wang B S, and  Sun Y P 2013
\emph{Chin. Phys. B} {\bf22}  067501 

\bibitem{Wiendlocha-InBSc3}  Wiendlocha B,  Tobola J, Kaprzyk S, 
 Fruchart D, and  Marcus J 2006 \emph{Phys. Status Solidi B} {\bf243}  351 

\bibitem{Wiendlocha-ABSc3} Wiendlocha B,  Tobola J, and  Kaprzyk S 2006
\emph{Phys. Rev. B} {\bf73}  134522  

\bibitem{Wiendlocha-ANCr3}  Wiendlocha B,  Tobola J,  Kaprzyk S,
and  Fruchart D 2007 \emph{J. Alloys Compd.} {\bf442}  289  

\bibitem{RhNCr3}  T\"{u}t\"{u}nc\"{u} H M and  Srivastava G P 2012 \emph{J. Appl. Phys.}
{\bf112} 093914 

\bibitem{GaNCr3-sc}  T\"{u}t\"{u}nc\"{u} H  M and  Srivastava G P 2013 \emph{J. Appl.
Phys.} {\bf114} 053905
\bibitem{blochl1994projector}  Bl{\"o}chl P  E  1994 \emph{Phys. Rev. B} {\bf50} 
17953 

\bibitem{torrent2008implementation}  Torrent M,  Jollet F,  Bottin F,
 Z{\'e}rah G, and  Gonze X 2008 \emph{Comp. Mater. Sci.} {\bf42}  337  

\bibitem{gonze2002firstprinciples}  Gonze X,  Beuken J M,  Caracas R,
 Detraux F,  Fuchs M,  Rignanese G M,  Sindic L,  Verstraete M, 
Zerah G,  Jollet F,  \emph{et al}. 2002 \emph{Comp. Mater. Sci.} {\bf25} 478 


\bibitem{gonze2009abinitfirstprinciples} Gonze X,  Amadon B,  Anglade P M,  Beuken J M,  Bottin F,  Boulanger P,  Bruneval F,  Caliste D,  Caracas R,  Cote M, \emph{et al}. 2009 \emph{Comput. Phys. Comm.} {\bf180}  2582 

\bibitem{gonze_brief_2005}  Gonze X,  Rignanese G M,  Verstraete M, Beuken  J M,  Pouillon Y,  Caracas R,  Jollet F,  Torrent M,  Zerah G,  Mikami M, \emph{et al}. 2005 \emph{Z. Kristallogr.} {\bf220} 
558  



\bibitem{monkhorst1976special}  Monkhorst  H  J  and Pack J  D 1976 \emph{Phys.
Rev. B} {\bf13}  5188 

\bibitem{DFPT}  Baroni S,  de Gironcoli S,  Dal Corso A, and  Giannozzi P 2001
\emph{Rev. Mod. Phys.} {\bf73}  515 

\bibitem{US} Vanderbilt D 1990 \emph{Phys. Rev. B} {\bf41}  7892  

\bibitem{QE} Giannozzi P, \emph{et al}.  2009  \emph{J. Phys.: Condens. Matter} {\bf21}  395502 

\bibitem{perdew1996generalized} Perdew J  P, Burke  K, and Ernzerhof M 1996
\emph{Phys. Rev. Lett.} {\bf77} 3865 

\bibitem{Shao-ACCr3} Shao D  F,  Lu W J,  Lin S,  Tong P, and Sun Y
P 2013 \emph{ Adv. Condens. Matter. Phys.} {\bf2013}  136274  
 
\bibitem{Ignatov}  Ignatov A  Y,  Savrasov S Y, and  Tyson T A 2003
\emph{Phys. Rev. B} {\bf68}  220504(R) 

\bibitem{Heid}  Heid R,  Renker B,  Schober H,  Adelmann P,  Ernst D,
and  Bohnen K P 2004 \emph{Phys. Rev. B} {\bf69}  092511  

\bibitem{Jha-PRB}  Jha P  K 2005 \emph{Phys. Rev. B} {\bf72}  214502  

\bibitem{HMTUTU} T\"{u}t\"{u}nc\"{u}  H  M 
and  Srivastava G  P 2006 \emph{J. Phys.: Condens. Matter} {\bf18}  11089 


\bibitem{walte-rosner} W\"{a}lte A,  Fuchs G,  M\"{u}ller K H,  Handstein A,
 Nenkov K,  Narozhnyi V N,  Drechsler S L,  Shulga S,  Schultz L,
and  Rosner H 2004 \emph{Phys. Rev. B} {\bf70}  174503  



\bibitem{Dolgov} Dolgov O  V,  Golubov A A,  Mazin I I and
Maksimov  E G 2008 \emph{J. Phys.: Condens. Matter} {\bf20}  434226 

\bibitem{mcmillan1968transition} McMillan  W  L  1968 \emph{Phys. Rev.} {\bf167} 
331 

\bibitem{Dynes}  Dynes R 1972 \emph{Solid State Commun.} {\bf10}  615  

\bibitem{Allen-dynes} Allen P B and Dynes R C 1975 \emph{Phys. Rev. B} {\bf12}  905 

\bibitem{Oritenzi-Boeri} Ortenzi L,  Biermann S, 
Andersen O  K, Mazin I I , and  Boeri L 2011 \emph{Phys. Rev. B} {\bf83}  100505(R) 


\bibitem{Schaak}  Schaak R E,  Avdeev  M,  Lee W L,  Lawes G, 
Zandbergen H W,  Jorgensen J D,  Ong N P,  Ramirez A P, Cava R J 2004 \emph{J.
Solid State Chem.} {\bf177}  1244  

\bibitem{mazin-hcpiron}   Mazin I I,  Papaconstantopoulos  D A, and  Mehl M J  2002
\emph{Phys. Rev. B} {\bf65}  100511(R)  

\bibitem{Bose-hcpiron} Bose S K,   Dolgov O V,   Kortus J,  Jepsen O, and  Andersen O K   2003 \emph{Phys. Rev. B} {\bf67}  214518  



\bibitem{Boeri-PRL2008}  Boeri L,  Dolgov O V, and  Golubov A A 2008
\emph{Phys. Rev. Lett.} {\bf101}  026403  

\bibitem{Boeri-Physica C2009} Boeri L,  Dolgov O V, and  Golubov A  A 2009
\emph{Physica C} {\bf469}  628  

\bibitem{Mazin-PRL2008} Mazin I I,  Singh D J,  Johannes M D,
and  Du M H 2008 \emph{Phys. Rev. Lett.} {\bf101}  057003 

\bibitem{Boeri-Mazin-PRB2008}  Boeri L,  Calandra M,  Mazin I I,
 Dolgov O V, and  Mauri F 2010 \emph{Phys. Rev. B} {\bf82}  020506(R)    

\end{thebibliography}
\end{document}